\documentclass[11pt, letterpaper]{article}

\usepackage{amsmath}
\usepackage{amssymb}
\usepackage{amsthm}
\usepackage{graphicx}
\usepackage{hyperref}
\usepackage{url}
\usepackage{algorithm}
\usepackage{algorithmic}
\usepackage{caption}
\usepackage{subcaption}
\usepackage{booktabs}
\usepackage{authblk}
\usepackage{appendix}
\usepackage[utf8]{inputenc}
\usepackage[T1]{fontenc}
\usepackage[english]{babel}
\usepackage{geometry}
\usepackage{xcolor}
\usepackage{fancyhdr}
\usepackage{microtype}
\usepackage{parskip}
\usepackage{tabularx}
\usepackage{enumitem}
\usepackage{multirow}

\hypersetup{
    colorlinks=true,
    linkcolor=blue,
    filecolor=magenta,      
    urlcolor=cyan,
    citecolor=red,
}

\definecolor{headercolor}{RGB}{0, 50, 100}
\title{Segment Anything Model (SAM) for Radiation Oncology}

\author[1]{Lian Zhang}
\author[2]{Zhengliang Liu}
\author[3]{Lu Zhang}
\author[2]{Zihao Wu}
\author[3]{Xiaowei Yu}
\author[1]{Jason Holmes}
\author[1]{Hongying Feng}
\author[2]{Haixing Dai}
\author[4]{Xiang Li}
\author[4]{Quanzheng Li}
\author[2]{Dajiang Zhu}
\author[2]{Tianming Liu}
\author[1]{Wei Liu}

\affil[1]{Department of Radiation Oncology, Mayo Clinic}
\affil[2]{School of Computing, University of Georgia}
\affil[3]{Department of Computer Science and Engineering, The University of Texas at Arlington}
\affil[4]{Department of Radiology, Massachusetts General Hospital and Harvard Medical School}

\date{}

\begin{document}

\maketitle

\begin{abstract}

In this study, we evaluate the performance of the Segment Anything Model (SAM) in clinical radiotherapy. We collected clinical cases from four disease sites at Mayo Clinic: prostate, lung, gastrointestinal, and head \& neck, which are major treatment sites in radiation oncology. For each case, we selected the organs-at-risk (OARs) in radiotherapy planning. We then compared both the Dice coefficients and Jaccard indices derived from three distinct methods: clinical manual delineation (considered to be the ground truth), automatic segmentation using SAM's 'segment anything' mode, and automatic segmentation using SAM's 'box prompt' mode. Our results indicate that SAM's segment anything mode can achieve clinically acceptable segmentation results in most OARs with Dice scores higher than 0.7. SAM's box prompt mode further improves the Dice scores by 0.1$\sim$0.5. The results show that SAM performs better in automatic segmentation for the prostate and lung sites, while its performance lags behind in the gastrointestinal and head \& neck sites. When considering the size of the organ and the clarity of its boundary, SAM displays better performance for large organs with clear boundaries, such as the lung and liver, and worse for smaller organs with unclear boundaries, like the parotid and cochlea. These findings align with the generally accepted variations in difficulty level associated with the manual delineation of different organs at different sites in clinical radiotherapy. Given that SAM, a model pre-trained purely based on natural images, could handle the delineation of OARs from medical images with clinically acceptable accuracy, these results highly demonstrate SAM's robust generalization capabilities with consistent accuracy in automatic segmentation for radiotherapy, i.e., achieving delineation of different OARs of different sites using a generic automatic segmentation model. SAM's generalization capabilities across different disease sites make it technically feasible to develop a generic model for automatic segmentation in radiotherapy. Further research to enhance SAM to support 3D segmentation and multiple modalities is warranted.

\end{abstract}

\section{Introduction}

Recent advancements in natural language processing (NLP) have led to large language models (LLMs) that can generalize to new domains with little training data \cite{zhao2023brain,liu2023summary}. Models such as GPT-3 \cite{brown2020language}, ChatGPT \cite{liu2023summary}, GPT-4 \cite{openai2023gpt}, and Google's PaLM-2 \cite{anil2023palm} have revolutionized NLP. LLMs enable artificial intelligence (AI) systems that can perform a wide range of language tasks \cite{holmes2023evaluating,liu2023radiology,ma2023impressiongpt, wu2023exploring,zhong2023chatabl,liu2023deid,dai2023chataug,liao2023differentiate} in diverse domains \cite{rezayi2023exploring,dai2023ad,liu2023radiology}. Their significant success has inspired interest in building similar "foundation models" for computer vision \cite{li2023artificial,huang2023segment,zhao2023brain}.

In response, Meta's Segment Anything Model (SAM) \cite{Kirillov23} was proposed as a generalized and promptable model for image segmentation. SAM is trained on over 1 billion masks in 11 million natural images and can generate segmentation masks for any object based on prompts. SAM shows strong performance on natural images and may enable zero-shot learning for new objects without retraining. SAM's capability suggests that it could enhance interactive medical image segmentation where physicians provide guidance to generate accurate delineations. While SAM's performance on natural images is impressive, medical images possess unique challenges. SAM was not designed specifically for medical images and may struggle with their complexity \cite{huang2023segment}.

Since SAM was proposed, various foundation models have emerged. In image editing, the inpaint anything (IA) framework \cite{yu2023inpaint} has successfully integrated SAM with state-of-the-art image inpainters \cite{suvorov2022resolution} and AI-generated content (AIGC) models \cite{rombach2022high}. This integration has resulted in a powerful pipeline capable of addressing various challenges in inpainting-related tasks. Another notable contribution is the "edit everything" approach \cite{xie2023edit}, which follows a similar pipeline to IA. In this case, SAM is utilized to segment the input image without any prompts, and subsequently, a source prompt is employed to guide CLIP (Contrastive Language–Image Pre-training) \cite{radford2021learning} for image editing using simple text instructions. Furthermore, SAM has also found applications in style transfer. The Any-to-Any Style Transfer \cite{liu2023any} leverages SAM’s promotable region selection capability to facilitate effective style transfer between different images. Additionally, SAM’s versatility extends to object detection tasks. Giannakis et al. \cite{giannakis2023deep} propose a universal crater detection scheme utilizing the zero-shot generalization of SAM, enabling the detection of unfamiliar objects. In addition to its various applications in natural images, SAM has been evaluated across diverse real-world segmentation scenarios \cite{ji2023segment} involving different types of images, such as medical images in the healthcare domain.

In recent years, an escalated incidence of cancer has been observed, with approximately half of these cases necessitating radiation therapy (RT), as indicated by evidence-based assessments \cite{delaney2005role}. Although efficacious in eradicating tumor cells, RT concurrently poses the risk of damaging nearby normal tissues, potentially leading to an array of complications. It is therefore of paramount importance to accurately delineate OARs proximal to the tumor on simulated computed tomography (CT) images prior to formulating the treatment plan. To minimize tissue damage, the dose to these OARs should be minimized as much as possible during the treatment planning, guided by the accurately delineated OARs \cite{schild2014proton,shan2018robust,zaghian2014automatic,zaghian2017comparison,liu2019system,liu2019dosimetric,liu2018small,liu2015robustness,liu2016robustness,liu2012robust}. 
Currently, the process of OAR delineation is manual and is executed by a radiation oncologist or a dosimetrist. It involves an in-depth analysis of CT images on a slice-by-slice basis, a procedure that is both labor-intensive and time-consuming \cite{harari2010emphasizing,ding2023deep}. The burgeoning OAR delineation demand for RT inevitably extends patients' waiting times, a factor shown to adversely impact tumor local control and prognosis due to tumor proliferation \cite{chen2008relationship}. For adaptive RT, which necessitates frequent delineation of OARs and target volumes on verification CT images along with alterations to treatment plans during the course of treatment, the need for expeditious and accurate delineation is even more pronounced. Furthermore, imprecise delineation may lead to sub-optimal treatment plans and unintended complications. Successful radiation therapy necessitates high geometric and dosimetric precision \cite{liu2015impact,liu2014dosimetric,liu2016exploratory,feng2020beam,shan2018robust}. The rising trend of pencil beam scanning proton therapy has afforded us the ability to more accurately shape dose distributions to align with the treatment target, whilst minimizing the exposure to OARs \cite{schild2014proton,an2017robust,liu2018small,liu2013effectiveness,li2015robust,liu2019treatment,liu2013ptv,liu2012influence}. Nonetheless, the sharp dose gradients facilitated by these methods could potentially pose a risk if the delineation of structures is imprecise.Therefore, the rapid and precise delineation of OARs becomes even more vital for proton therapy, especially in regions where there is a dearth of experienced radiation oncologists.

An automatic segmentation (AS) algorithm can potentially address these challenges, thereby considerably enhancing RT efficiency \cite{chen2021deep}. Numerous auto-segmentation methodologies have been proposed, including deformable image registration (DIR), atlas-based auto-segmentation, and the more recent deep learning-based segmentation (DLS) \cite{kosmin2019rapid}. Although both DIR and atlas-based auto-segmentation have seen extensive implementation, their clinical utility is compromised due to limitations in accuracy and efficiency \cite{van2020improving,lee2019clinical}. The focus of auto-segmentation research has recently shifted towards AI methods, with a particular emphasis on those underpinned by deep learning (DL) \cite{chen2021deep,lin2021deep}. Over the past few years, the number of studies and clinical applications exploring DL-based segmentation of OARs in RT has proliferated, significantly improving the efficiency of auto-delineation in RT processes \cite{ahn2019comparative,elguindi2019deep,zhu2019anatomynet}. However, clinical feedback suggests a prominent issue with the current models: a lack of generalizability \cite{jarrett2019applications,bashyam2022deep}. These models often necessitate the training of a unique model for each RT site of one specific institution with medical images generated by a certain imaging protocol based on one specific imaging machine. Given the variety of RT sites, imaging protocols, and machines used, a multitude of independent auto-segmentation models need to be trained even within one institution, each requiring a considerable investment of time and a large volume of site/protocol/machine-specific training data. Some data harmonization methods have been proposed to mitigate this generalization problem with some limited success \cite{10.1007/978-3-031-16449-1_73}. In practical clinical scenarios, the determination of the site to be delineated often requires the invocation of a site/protocol/machine-specific model. The complexity inherent in such scenarios, including mixed-site images and images from rare sites, can significantly degrade the precision of existing auto-segmentation models, and in some cases, cause errors. Thus, the development of an universal auto-delineation model, capable of handling multiple sites simultaneously, could significantly augment the precision and efficiency of auto-segmentation in RT processes, particularly in delineating OARs.
An ideal solution would rely on limited data, generalize across protocols, modalities, anatomies, and institutions, and minimize human efforts. SAM demonstrates such potential but its performance on medical images remains unclear, especially for radiotherapy cases \cite{putz2023segment,huang2023segment}.

We evaluate SAM's ability to perform zero-shot segmentation of medical images from multiple anatomical sites from clinical radiation oncology with CT images (prostate, lung, gastrointestinal, and head\&neck). We assess SAM in "segment everything" mode where it generates masks for all objects and "box prompt" mode where users indicate regions of interest. Dice coefficient and Jaccard index are used to measure the spatial overlap between SAM's predictions and ground truth clinical delineations.

Our results provide a broad analysis of SAM for radiation oncology. SAM shows promise but performance varies significantly based on the disease sites and modalities. SAM excels at segmenting large, well-defined organs given unambiguous prompts but struggles with complex anatomies with ambiguous prompts, especially organs with unclear boundaries. Performance metrics suggest general trends but fuller qualitative analysis is warranted to determine practical utility, and there is much room for adapting and improving SAM for radiation oncology applications. 

Our key contributions are:
\begin{itemize}

\item We provide critical insights into the capabilities and constraints of SAM in the context of radiation oncology.

\item We illuminate the necessity for stringent analysis and adaptation of foundation models like SAM for specialized areas prior to their deployment in clinical settings.

\item We elaborate on how judicious application of SAM, in collaboration with human expertise, can expedite and refine the task of medical image segmentation.

\item We advocate for a balanced view on the utility of foundation models in healthcare transformation, emphasizing a collaboration between AI and medical professionals.

\end{itemize}

\section{Related work}
\subsection{Segment Anything Model}
The Segment Anything project proposed by Meta is a groundbreaking initiative that aims to democratize the world of image segmentation, a vital task in computer vision \cite{Kirillov23}. It consists of two key components: a substantial dataset for image segmentation, and the SAM, a promtable foundation model. Taking inspiration from the world of NLP, the project creates a vast dataset and a segmentation model, both open-sourced, thus opening up enormous possibilities. SAM, which can run in real-time in browsers, offers a highly automated image segmentation approach that requires minimal human intervention. It's a deep learning model, trained on over 1 billion masks in 11 million images, which can cut out almost anything from an image. Unlike traditional models that require specialized training, SAM is generalizable and can respond to user prompts about specific areas to segment.

This model includes three essential components: an image encoder, a prompt encoder, and a mask decoder. Input images pass through an image encoder to produce an embedding, and the model can accept prompts as points, boxes, or rough masks. For more nuanced prompting, the authors are working on a version of SAM that accepts text input, similar to language models. The resultant segmented image comes with multiple valid masks and a confidence score, signifying the segmentation's accuracy. SAM's capabilities could revolutionize various fields, from augmenting reality through precise object identification to biomedical image segmentation, aiding cell microscopy analysis \cite{Mazurowski23}. It can also be integrated with diffusion-based image generation models for efficient image inpainting and be utilized for generating semantic segmentation datasets, making it an exciting development in the realm of AI and computer vision \cite{Yu23SAM}.

\subsection{SAM for Medical Imaging}
Since SAM was proposed as an innovative framework for image segmentation, its application within the realm of medical imaging has increasingly become a topic of interest within the field \cite{huang2023segment} \cite{JiWei23SAM}. SAM has been tested under the 'everything' mode for segmenting lesion regions across a range of anatomical structures, such as the brain, lung, and liver, and imaging modalities like CT and Magnetic Resonance Imaging (MRI). The experimental findings suggest that while SAM exhibits relative proficiency in segmenting organs with distinct boundaries, it encounters difficulties when trying to accurately identify lesions with irregular shapes and contours \cite{Ji23SAM}. Another study drew comparisons between the SAM and the Brain Extraction Tool (BET) from the FMRIB Software Library in brain extraction tasks, and the quantitative analysis revealed superior segmentation performance by SAM compared to BET, thus illustrating SAM's promising potential for use in brain segmentation tasks \cite{Mohapatra23SAM}. SAM was also applied to digital pathology segmentation tasks, which encompassed the segmentation of tumor and non-tumor tissue, as well as cell nuclei, on whole-slide images. The findings indicated that SAM exhibits excellent performance when segmenting large, interconnected objects. However, its reliability wavers when tasked with dense instance object segmentation, even when provided with all the target boxes \cite{Deng23SAM}. SAM for polyp segmentation tasks using five benchmark datasets, all under the 'everything' setting has been reported. The findings indicate that while SAM has the capacity to accurately segment polyps in certain instances, it significantly lags behind the performance of leading-edge methods, suggesting substantial room for improvement \cite{Zhou23SAM}. Extensive experimentation on multiple public datasets shows that the zero-shot segmentation capabilities of SAM are not sufficiently robust for direct application in medical image segmentation \cite{huang2023segment} \cite{He23SAM} \cite{Mazurowski23}. 

\subsection{Cancer Segmentation}
The field of cancer segmentation in medical imaging has seen considerable advancements in recent years, motivated by the potential to improve the accuracy of cancer detection, diagnosis, and radiotherapy. The goal of cancer segmentation is to accurately delineate the boundaries of cancerous regions within medical images.

Initial works on cancer segmentation relied heavily on traditional image processing techniques \cite{Vese02}. These included thresholding, edge detection, region-growing, and clustering-based techniques, among others. An example of these methods includes the level set method, which has been widely used in brain tumor segmentation \cite{Chan01}. The advent of machine learning (ML) brought significant improvements to cancer segmentation \cite{Zhang01}. Techniques such as support vector machines (SVM), decision trees, and random forests were utilized to improve the segmentation process. Furthermore, hand-crafted features were used to capture the characteristics of cancerous tissue, contributing to the performance of ML-based models.

The recent explosion of deep learning techniques has revolutionized the field of cancer segmentation \cite{Litjens17}. For instance, U-Net \cite{Unet2015}, a type of convolutional neural networks (a widely used architecture in medical image analysis \cite{ding2022accurate,bi2023community,liu2022discovering,dai2022graph,qiang4309357deep}), has been applied to diverse tasks \cite{siddique2021u,zhang2023beam,ding2023deep}, including biomedical image segmentation such as lung cancer segmentation from CT scans and brain tumor segmentation from MRI images \cite{Unet2015,liu2022survey,siddique2021u}. Developing fully automated systems for cancer segmentation is a trend in recent research \cite{Ardila19}. These systems aim to minimize the need for manual intervention, thus reducing the time taken for diagnosis and the potential for human error. 

The use of SAM for cancer segmentation has seen a significant increase since it was proposed. With point or bounding box prompts, SAM achieves competitive results comparable to supervised training models like U-net in the areas of liver tumor, breast tumor, and colon polyp segmentation \cite{hu2023sam,hu2023breastsam,li2023polyp}. By fine-tuning the image encoder, prompt encoder, and mask decoder using the same labeled data, SAM can even outperform state-of-the-art results in skin cancer segmentation \cite{hu2023skinsam}. These studies demonstrate that the pre-trained SAM is efficient and effective for computer-aided rapid cancer detection and diagnosis. Furthermore, it can also serve as a versatile foundation model for developing domain-specific models fine-tuned for targeted cancer segmentation.

Despite significant progress, challenges remain in the field of cancer segmentation. These include dealing with the high variability in cancer appearance, the scarcity of annotated medical images, and the need for reliable evaluation metrics. Recent research has started exploring the potential of advanced deep learning techniques, including generative models \cite{rouzrokh2022multitask,chen2018deep} and self-supervised learning \cite{shurrab2022self}, to address these challenges.

\section{Methodology}
\subsection{Datasets}
\begin{figure}
    \centering
    \includegraphics[width=1\linewidth]{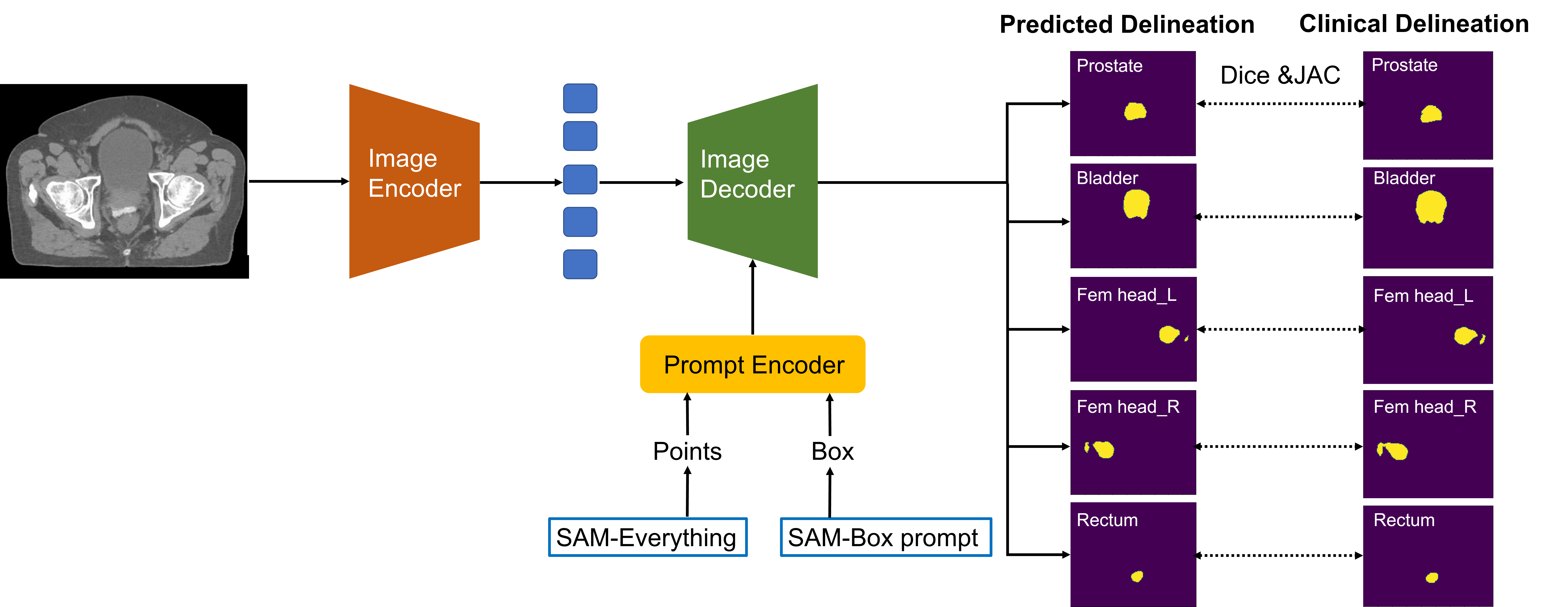}
    \caption{Workflow of SAM segmentation in clinical cases with SAM segment everything mode and SAM box prompt mode taking prostate as an example.}
    \label{fig:1}
\end{figure}

This study has received approval from the Institutional Review Board (IRB) at Mayo Clinic. Figure \ref{fig:1} illustrates the overall framework of our work to set up SAM for clinical case segmentation. Algorithm \ref{algorithm_1} further illustrates the internal algorithm process of SAM. We collected case images from the four most common sites in clinical radiotherapy, namely the prostate, gastrointestinal, lungs, and head \& neck. We gathered 20 patients from each site, totaling 80 patients. From the perspective of clinical radiotherapy delineation, we divided the cases into two groups: the simple group includes the prostate and lungs, while the difficult group encompasses the gastrointestinal and head \& neck. As SAM currently only supports 2D delineation, for a fair comparison, we evaluated the delineation of five typical 2D slices extracted from each patient that best represented the anatomical information. Based on the OARs that require attention during the formulation of the radiotherapy plan, we selected regions of interest for delineation comparison for different disease sites, following the recommendations of the Radiation Therapy Oncology Group (RTOG).

For the prostate, we selected the following regions of interest for evaluation: prostate, bladder, left femoral head, right femoral head, and rectum. For the lungs, we evaluated the following areas: left lung, right lung, heart, spinal cord, and esophagus. For the gastrointestinal, we assessed the liver, stomach, left kidney, right kidney, spinal cord, large bowel, and small bowel. For the head \& neck, we evaluated the brain, left parotid, right parotid, spinal cord, mandible, left cochlea, and right cochlea. All manual delineations were performed by highly experienced radiation oncologists and meet the RTOG delineation standards, and are used in the formulation of clinical radiotherapy plans.

\begin{algorithm}[tb]
\caption{The SAM Inference Process.}
\label{algorithm_1}
\textbf{Input}: Checkpoint $checkpoint$, input image $input\_image$, prompt $prompt$ \\
\textbf{Output}: Mask $mask$
\begin{algorithmic}
    \STATE {\bfseries Define Function InitializeModel:} \\
    \STATE \quad $model \gets$ SAM model built from $checkpoint$
    \STATE
    \STATE {\bfseries Define Function ImagePreparation:} \\
    \STATE \quad $processed\_image \gets$ normalized and standardized version of $input\_image$
    \STATE
    \STATE {\bfseries Define Function InputEncoding:} \\
    \STATE \quad $image\_embedding \gets$ image encoder of $model$ applied on $processed\_image$ \\
    \STATE \quad $prompt\_embedding \gets$ prompt encoder of $model$ applied on $prompt$
    \STATE
    \STATE {\bfseries Define Function MaskCreation:} \\
    \STATE \quad $mask \gets$ mask decoder of $model$ applied on $image\_embedding$ and $prompt\_embedding$
    \STATE
    \STATE $model$ = InitializeModel($checkpoint$)
    \STATE $processed\_image$ = ImagePreparation($input\_image$)
    \STATE $image\_embedding$, $prompt\_embedding$ = InputEncoding($model$, $processed\_image$, $prompt$)
    \STATE $mask$ = MaskCreation($model$, $image\_embedding$, $prompt\_embedding$)
    \STATE return $mask$
\end{algorithmic}
\end{algorithm}

\subsection{Automatic segmentation of everything}
In the 'segment everything' mode, SAM is designed to create segmentation masks for every possible object present within the full image, no manual priors are needed. This mode is considered the first testing approach. The commencement of this method involves producing a grid of point prompts, also known as grid sampling, which spans the entire image. To enhance the segmentation of the target regions, more random point prompts will be assigned to the target regions guiding an improved segmentation process in this study. Following that, the prompt encoder uses the sampled grid points to generate point embeddings, which are then merged with the image embeddings. The mask decoder then receives this blend as input and delivers multiple potential masks for the entire image. Afterward, a filter system is put into action to eliminate duplicate and inferior masks.
\subsection{Manual box prompt}
In the prompt mode, the box prompt signifies the spatial region that necessitates segmentation, representing the object of interest. This principal mode of evaluation employs prompts that mimic a human user's interaction, crafted while closely observing the objects. Our focus lies primarily on the box prompt, tailored to encapsulate SAM's realistic use cases for creating image masks. An experienced medical physicist typically places the box prompt, guided by anatomical characteristics and clinical experience. The placement is usually in close proximity to the region of interest margin. It's vital to remember that a single "object" of interest or a "ground truth" mask might comprise of multiple disconnected segments, a situation commonly encountered in medical imaging. To ensure each distinct, contiguous region of the object of interest is accurately represented, multi-box prompts are strategically placed.
\subsection{Evaluation metrics}
In order to thoroughly assess SAM's segmentation performance, we employed two commonly used metrics, as detailed below:

1) Dice Coefficient (DICE, \%): This measure of similarity is used to evaluate the degree of overlap between the prediction and the ground truth (GT) as defined as Eq. \ref{eq:1}. With a range between [0, 1], a higher value denotes a more successful performance by the model.

\begin{equation}
\label{eq:1}
 DICE\left(y,\widetilde{y}\right){=\frac{2\left|y\cap\widetilde{y}\right|}{\left|y\right|+\left|\widetilde{y}\right|}}   
\end{equation}

2) Jaccard Index (JAC, \%): Also recognized as the Intersection over Union (IOU), this metric, although similar to DICE, poses more stringent demands as defined as Eq. \ref{eq:2}. It quantifies the complete overlap of label ensembles across multiple test images, accommodating fractional labels through the application of fuzzy set theory. Like the DICE coefficient, higher JAC values signify superior model performance.

\begin{equation}
\label{eq:2}
JAC\left(y,\widetilde{y}\right)=\frac{\left|y\cap\widetilde{y}\right|}{\left|y\cup\widetilde{y}\right|}
\end{equation}

where \textit{y} denotes the volume of clinical manual delineation, and \(\widetilde{y}\) denotes the volume of SAM auto-segmentation.

\section{Results}
\subsection{Example cases}
\begin{figure}
    \centering
    \includegraphics[width=1\linewidth]{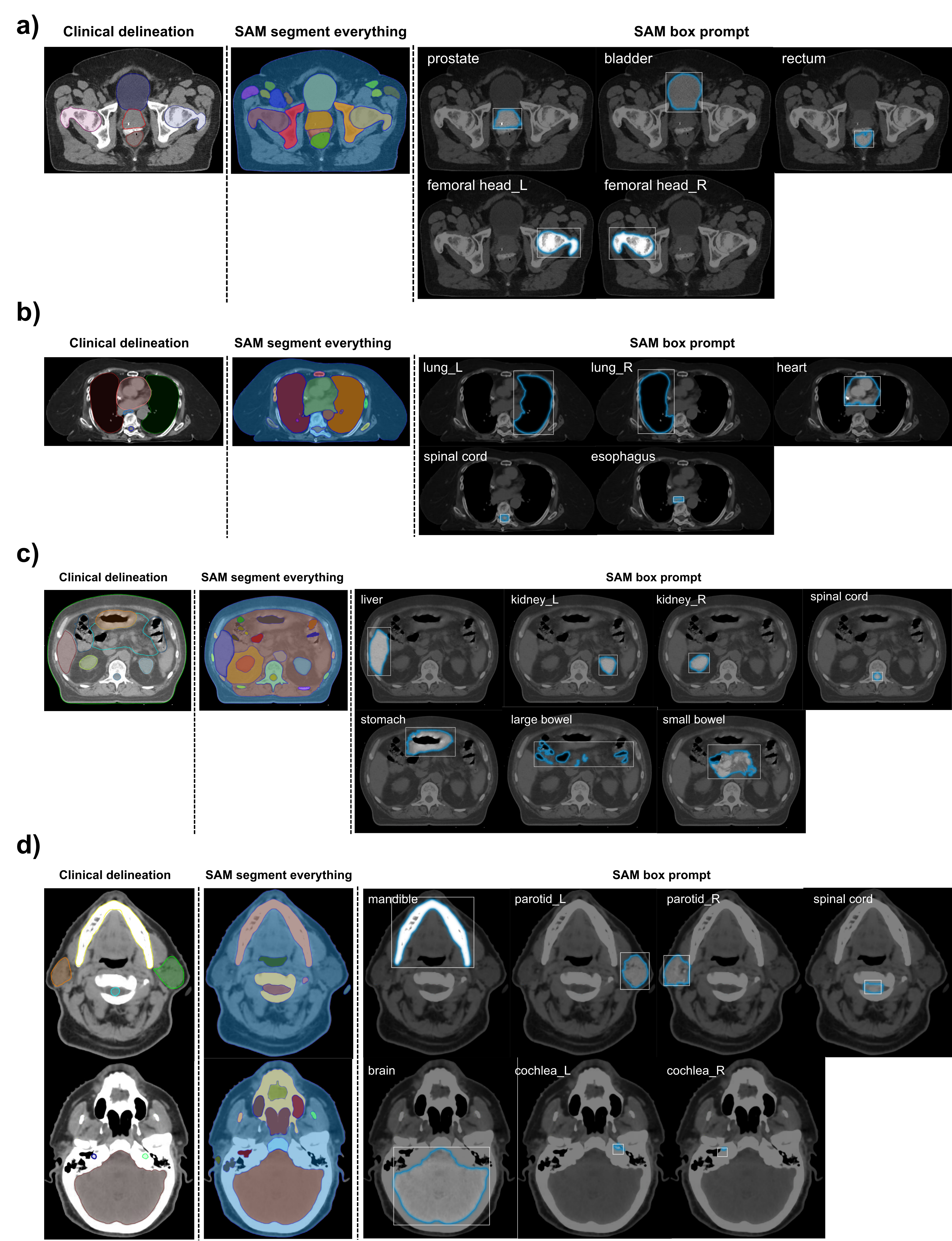}
    \caption{Comparison of the clinical delineation (ground truth), SAM segment everything, and SAM with manual box prompt of example cases. a) prostate  b) lung  c) gastrointestinal d) head\&neck}
    \label{fig:2}
\end{figure}

Figure \ref{fig:2} shows the auto-segmented contours from the two experiments (SAM segment everything and SAM box prompt) and the clinical delineation in the axial plane of one typical prostate (Fig. \ref{fig:2}(a)) case, one typical lung (Fig. \ref{fig:2}(b)) case, one typical gastrointestinal (Fig. \ref{fig:2}(c)) case, and one typical head \& neck (Fig. \ref{fig:2}(d)) case.
\subsection{SAM model accuracy}
For the prostate, as shown in Figure \ref{fig:3} and Figure \ref{fig:5}, the SAM segment everything mode resulted in a Dice score of 0.709±0.043 and a JAC (Jaccard Index) score of 0.551±0.053 when outlining the prostate. In the bladder adjacent to the prostate, the Dice score was 0.748±0.036 and the JAC score was 0.598±0.046. For the femoral head\_L and femoral head\_R, their Dice scores were both around 0.8, and JAC scores were around 0.7. The Dice score for the rectum was relatively low, at 0.634±0.045, and the JAC score was 0.466±0.051.

For the lungs, as shown in Figure \ref{fig:3} and Figure \ref{fig:5}, SAM segment everything mode produced Dice scores around 0.86 for both lung\_L and lung\_R, and JAC scores around 0.75. The heart had a Dice score of 0.671±0.031 and a JAC score of 0.506±0.032. The spinal cord had a relatively low Dice score of 0.457±0.038, and a JAC score of 0.297±0.029. For the esophagus, SAM segment everything mode was unable to outline or recognize it, resulting in Dice and JAC scores of around 0.

For the gastrointestinal, as shown in Figure \ref{fig:4} and Figure \ref{fig:6}, SAM segment everything mode produced a Dice score of 0.856±0.021 and a JAC score of 0.750±0.032 when outlining the liver. The results for kidney\_L and kidney\_R were similar, with Dice scores around 0.8 and JAC scores around 0.7. The spinal cord results were consistent with those of the lung, with Dice and JAC scores of about 0.45 and 0.3 respectively. The Dice score for the stomach was relatively low, at 0.319±0.027, and the JAC score was 0.190±0.021. For the large bowel and small bowel, SAM segment everything mode was unable to outline or recognize them, resulting in Dice and JAC scores of around 0.

For the head \& neck, as shown in Figure \ref{fig:4} and Figure \ref{fig:6}, SAM segment everything mode produced a Dice score of 0.896±0.018 and a JAC score of 0.812±0.030 when outlining the brain. The mandible had a Dice score of 0.868±0.022, and a JAC score of 0.768±0.033. The spinal cord had similar results to other sites, with Dice and JAC scores of about 0.4 and 0.3 respectively. For the parotid\_L, parotid\_R, cochlea\_L, and cochlea\_R, SAM segment everything mode was unable to outline or recognize them, resulting in Dice and JAC scores of around 0. All results were summarized in Table \ref{table:1} (the third and fifth column).

\subsection{Impact of box prompt}
Overall, after introducing the manual box prompt, there was an improvement in Dice and JAC results for most organs. Some organs that could not be identified in SAM segment everything mode were now recognizable, although the resulting Dice and JAC scores were low. However, a few small organs in the head \& neck remained unrecognizable even after the box prompt was employed.

\begin{figure}
    \centering
    \includegraphics[width=0.9\linewidth]{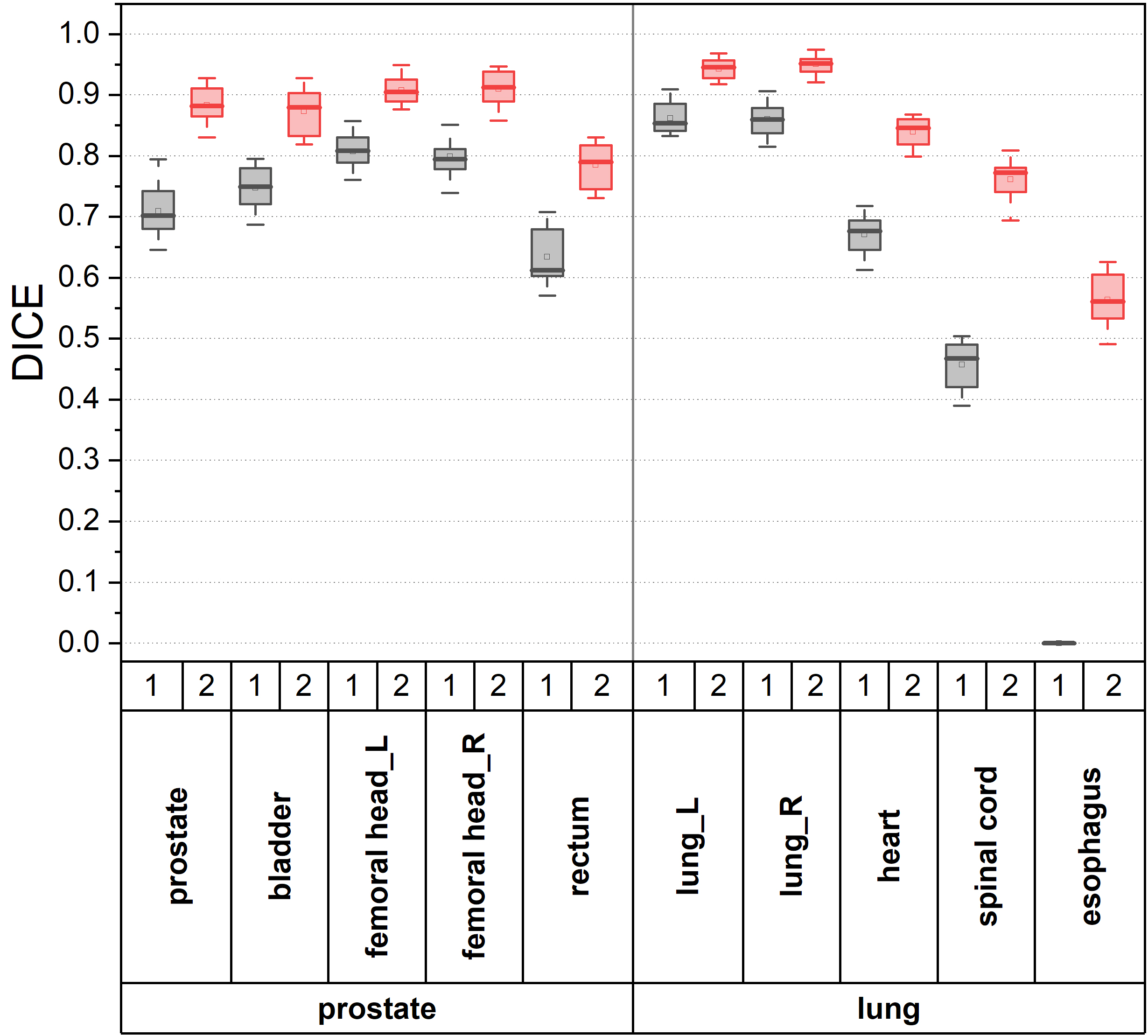}
    \caption{Boxplot (minimum, first quartile, median, third quartile, and maximum, respectively) of Dice coefficients of OARs between the ground clinical delineation and the SAM auto-segmented ones from three different experiments of the testing cases for prostate and lung. SAM segment everything and SAM manual box prompt correspond to 1 (grey) and 2 (red) in the figure.}
    \label{fig:3}
\end{figure}

\begin{figure}
    \centering
    \includegraphics[width=0.9\linewidth]{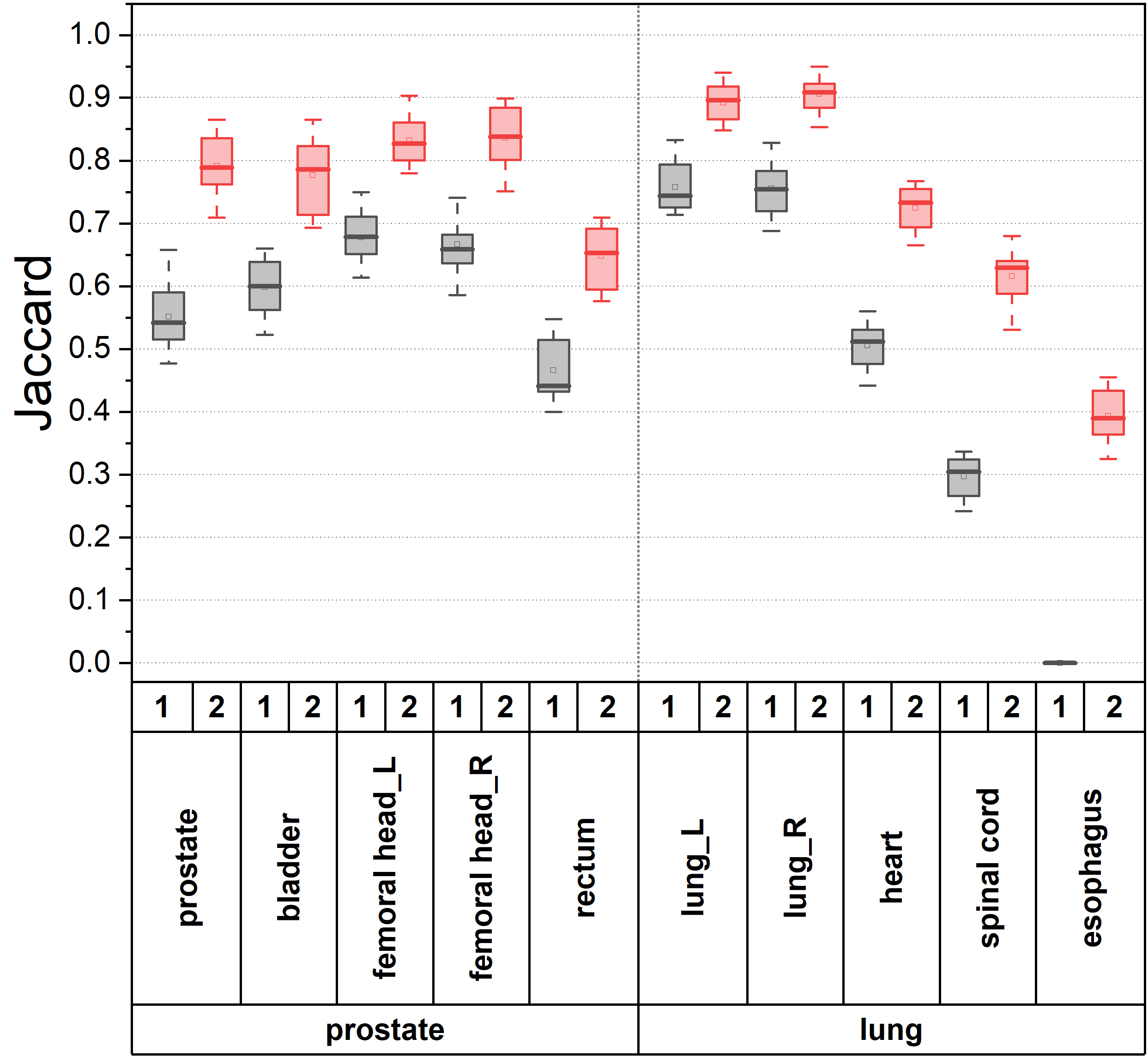}
    \caption{Boxplot (minimum, first quartile, median, third quartile, and maximum, respectively) of Jaccard coefficients of OARs between the ground clinical delineation and the SAM auto-segmented ones from three different experiments of the testing cases for prostate and lung. SAM segment everything and SAM manual box prompt correspond to 1 (grey) and 2 (red) in the figure.}
    \label{fig:5}
\end{figure}

For the prostate, as shown in Figure \ref{fig:3} and Figure \ref{fig:5}, the manual box prompt mode resulted in a Dice score of 0.883±0.030 and a JAC score of 0.792±0.049 when outlining the prostate. The bladder had a Dice score of 0.873±0.037, and a JAC score of 0.776±0.059. For femoral head\_L and femoral head\_R, their Dice scores each increased by around 0.1, and their JAC scores each increased by about 0.17. The rectum's Dice score increased to 0.785±0.031, and the JAC score to 0.648±0.047.

For the lungs, as depicted in Figure \ref{fig:3} and Figure \ref{fig:5}, manual box prompt mode improved the Dice scores for both lung\_L and lung\_R by about 0.09, and the JAC scores by about 0.15. The spinal cord had a Dice score of 0.760±0.032, and a JAC score of 0.615±0.047-. The heart had an improvement with Dice scores and JAC scores increasing by about 0.16 and 0.21, respectively. The esophagus was recognizable under manual box prompt mode, but both Dice and JAC scores were low, at about 0.56 and 0.39, respectively.

\begin{figure}
    \centering
    \includegraphics[width=0.9\linewidth]{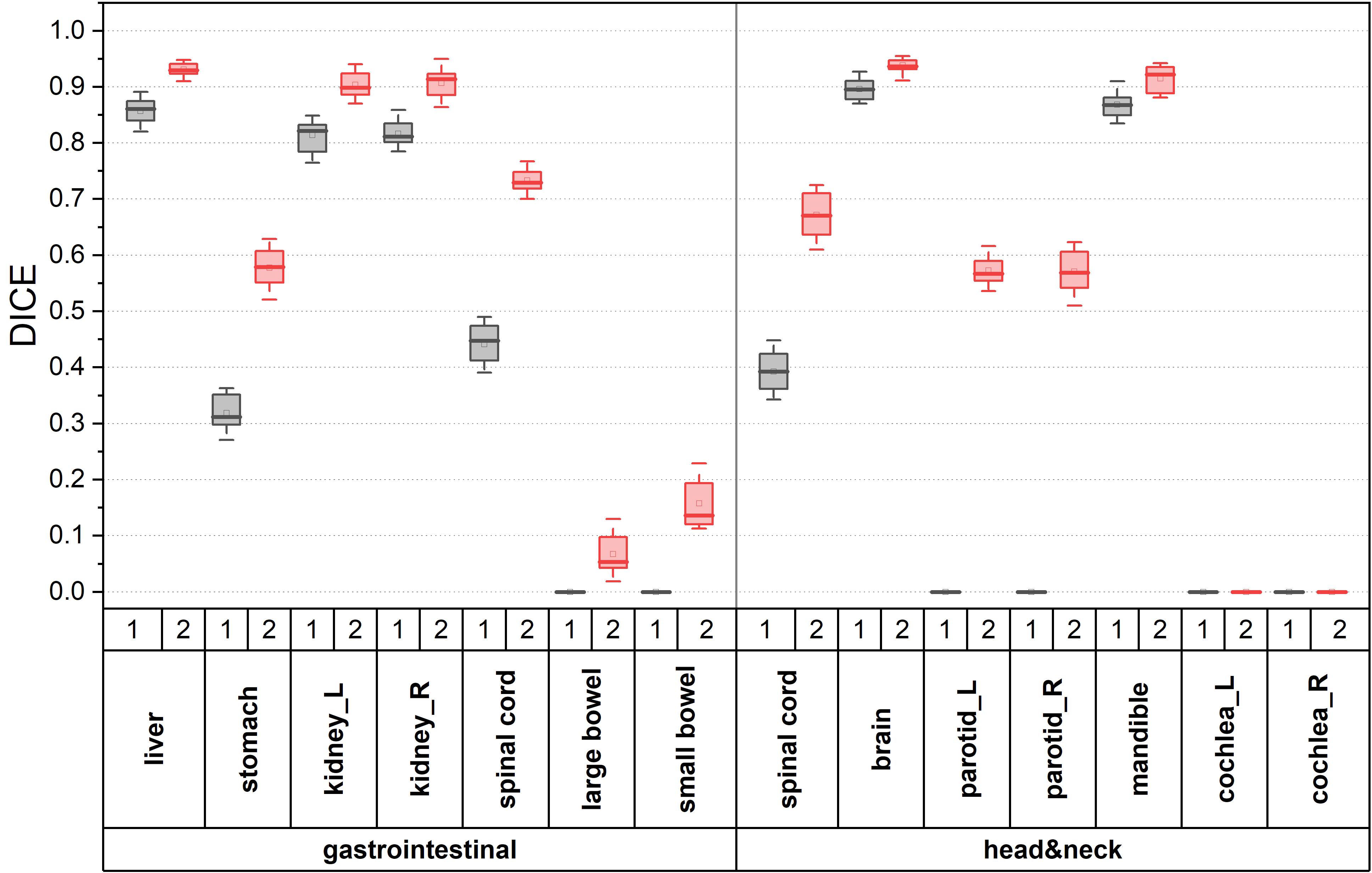}
    \caption{Boxplot (minimum, first quartile, median, third quartile, and maximum, respectively) of Dice coefficients of OARs between the ground clinical delineation and the SAM auto-segmented ones from three different experiments of the testing cases for gastrointestinal and head\&neck. SAM segment everything and SAM manual box prompt correspond to 1 (grey) and 2 (red) in the figure.}
    \label{fig:4}
\end{figure}

\begin{figure}
    \centering
    \includegraphics[width=0.9\linewidth]{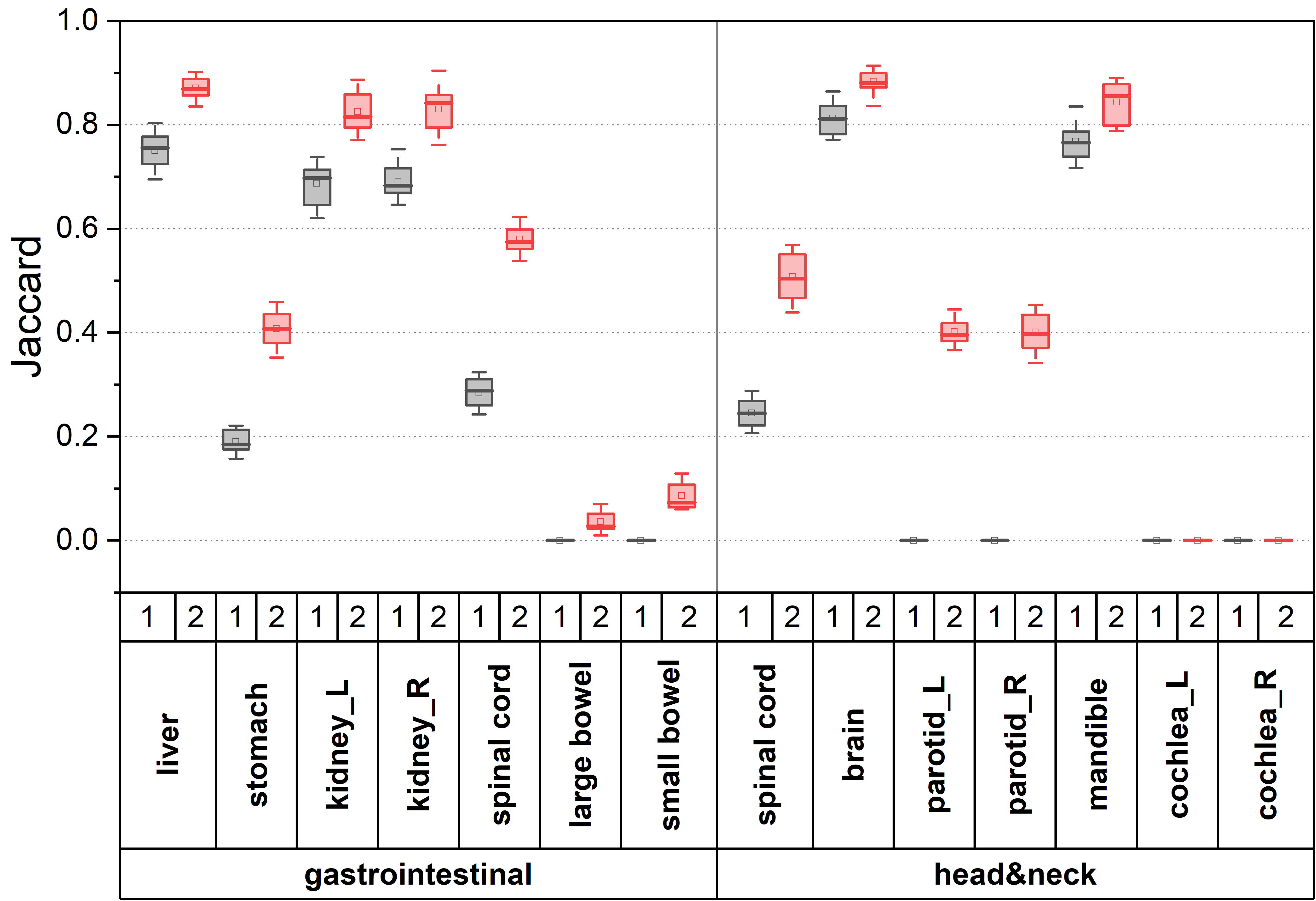}
    \caption{Boxplot (minimum, first quartile, median, third quartile, and maximum, respectively) of Jaccard coefficients of OARs between the ground clinical delineation and the SAM auto-segmented ones from three different experiments of the testing cases for gastrointestinal and head\&neck. SAM segment everything and SAM manual box prompt correspond to 1 (grey) and 2 (red) in the figure.}
    \label{fig:6}
\end{figure}

For the gastrointestinal, as shown in Figure \ref{fig:4} and Figure \ref{fig:6}, the manual box prompt mode yielded a Dice score of 0.930±0.011 and a JAC score of 0.871±0.019 when outlining the liver. For kidney\_L and kidney\_R, the Dice and JAC scores both increased by about 0.09 and 0.13 respectively. The spinal cord saw an increase of around 0.3 for both the Dice and JAC scores. For the stomach, the Dice and JAC scores increased by about 0.26 and  0.21 respectively. The large bowel was recognizable under manual box prompt mode, but both Dice and JAC scores were low, at around 0.07 and 0.04 respectively. For the small bowel, manual box prompt mode recognized it, with both Dice and JAC values being around 0.15 and 0.09 respectively.

For the head \& neck, as shown in Figure \ref{fig:4} and Figure \ref{fig:6}, the manual box prompt mode resulted in a Dice score of 0.938±0.012 and a JAC score of 0.883±0.021 when outlining the brain. The mandible had a Dice score of 0.915±0.021 and a JAC score of 0.843±0.039. The spinal cord had similar improvements as other regions, with increases of around 0.29 and 0.26 for the Dice and JAC scores respectively. For parotid\_L and parotid\_R, they were recognizable under manual box prompt mode, but both Dice and JAC scores were low, at around 0.57 and 0.40 respectively. For cochlea\_L and cochlea\_R, manual box prompt mode was still unable to recognize them, with both Dice and JAC values being around 0. All results were summarized in Table \ref{table:1} (the fourth and sixth column).


\begin{table}
\label{table:1}
\centering
\caption{Dice and Jaccard scores of reference ROIs segmented by SAM in the prostate, lung,  gastrointestinal(GI), and head\&neck sites}
\resizebox{\linewidth}{!}{
\begin{tabular}{llllll}
\hline
\multirow{2}{*}{\textbf{Case Sites}} & \multirow{2}{*}{\textbf{Reference ROIs}} & \multicolumn{2}{c}{\textbf{Dice}}     & \multicolumn{2}{c}{\textbf{Jaccard}}  \\ \cline{3-6} 
                            &                                 & \textbf{SAM-Everything} & \textbf{SAM-Box}     & \textbf{SAM-Everything} & \textbf{SAM-Box}     \\ \hline
\multirow{5}{*}{\textbf{Prostate}}   & Prostate                        & 0.709±0.043    & 0.883±0.030 & 0.551±0.053    & 0.792±0.049 \\
& Bladder   & 0.748±0.036   & 0.873±0.037 & 0.598±0.046    & 0.776±0.059 \\
& Femoral head\_L                 & 0.807±0.028    &0.908±0.023 & 0.678±0.040 & 0.832±0.038 \\
  & Femoral head\_R                 & 0.799±0.038   & 0.910±0.029  & 0.669±0.056    & 0.836±0.048 \\
 & Rectum                          & 0.634±0.045    & 0.785±0.031 & 0.466±0.051   & 0.648±0.047 \\ \hline
\multirow{5}{*}{\textbf{Lung} }      & Lung\_L                         & 0.862±0.025    & 0.943±0.016 & 0.758±0.039     & 0.892±0.028 \\
& Lung\_R                         & 0.860±0.027   & 0.951±0.015  & 0.756±0.041    & 0.906±0.027 \\
& Heart                           & 0.671±0.031    & 0.839±0.023  & 0.506±0.032    & 0.723±0.035 \\
& Spinal cord                     & 0.457±0.038    & 0.760±0.032  & 0.297±0.029     & 0.615±0.047 \\
& Esophagus                       & 0    & 0.563±0.041 & 0    & 0.393±0.040 \\ \hline
\multirow{7}{*}{\textbf{GI} }        & Liver                           & 0.856±0.021     & 0.930±0.011 & 0.750±0.032    & 0.871±0.019 \\
& Kidney\_L                       & 0.813±0.029    & 0.904±0.023  & 0.687±0.039     & 0.826±0.037\\
& Kidney\_R                       & 0.817±0.022   & 0.906±0.026  & 0.691±0.031   & 0.829±0.043 \\
& Spinal cord                     & 0.441±0.039     & 0.733±0.020  & 0.283±0.027   & 0.579±0.028 \\
 & Stomach                         & 0.319±0.027    & 0.578±0.032  &0.190±0.021    & 0.407±0.030 \\
 & Small bowel                     & 0    & 0.159±0.041 & 0  & 0.086±0.026 \\
 & Large bowel                     & 0   & 0.068±0.036 & 0   & 0.036±0.023 \\ \hline
\multirow{7}{*}{\textbf{Head\&Neck} }                   & Brain                           & 0.896±0.018    & 0.938±0.012  & 0.812±0.030     & 0.883±0.021\\ & Mandible                        & 0.868±0.022    & 0.915±0.021 & 0.768±0.033     & 0.843±0.039 \\
& Spinal cord                     & 0.393±0.032     & 0.672±0.040 & 0.243±0.027    & 0.507±0.047 \\
& Parotid\_L                      & 0   & 0.573±0.026 & 0   & 0.401±0.023 \\
& Parotid\_R                      & 0    & 0.570±0.036 & 0    & 0.399±0.031 \\
& Cochlea\_L                      & 0   & 0 & 0  & 0 \\
& Cochlea\_R                      & 0& 0& 0    & 0 \\ \hline
\end{tabular}
}
\end{table}

\section{Discussion}

In this section, we analyze the current performance of SAM segmentation in radiation oncology and expand on the possible future perspectives, practical applications, and prospective improvements of SAM in the realm of radiation oncology, based on the detailed results and insights from our study.
\subsection{Current performance in radiation oncology-related segmentation}
As for the Dice and Jaccard results, under the SAM "segment everything" mode, the auto segmentation outcomes were satisfactory for the prostate's prostate, bladder, femoral head\_L, and femoral head\_R with (DICE: 0.7$\sim$0.8, JAC: 0.5$\sim$0.7), while they were less desirable for the rectum (DICE: $\sim$0.6, JAC: $\sim$0.4). For the lung, auto-segmentation for the lung\_L and lung\_R (DICE: 0.8$\sim$0.9, JAC: 0.7$\sim$0.8) were relatively better, yet less favorable for the heart (DICE: $\sim$0.6, JAC: $\sim$0.5) and spinal cord (DICE: $\sim$0.5, JAC: $\sim$0.3), with the esophagus not being recognized. The auto-segmentation outcomes for the gastrointestinal's liver, kidney\_L and kidney\_R (DICE: 0.8$\sim$0.9, JAC: 0.7$\sim$0.8) were relatively better, whereas those for the stomach (DICE: $\sim$0.3, JAC: $\sim$0.2) were less satisfactory, with the large bowel and small bowel not recognized. For the head \& neck, the brain and mandible (DICE: 0.8$\sim$0.9, JAC: 0.7$\sim$0.8)were better segmented, whereas the parotid\_L, parotid\_R, cochlea\_L, and cochlea\_R were not recognized. When comparing the results of segmentation across different sites, the "segment everything" mode in SAM performs better for the prostate and lung, but less satisfactory for the gastrointestinal and head \& neck. If the volume and clarity of an organ are taken into account, it can be observed that the model performs better at delineating organs with clear boundaries and larger volumes, such as the liver and brain, and less satisfactory for organs with indistinct boundaries and smaller volumes, such as the parotid and cochlea, which is in general agreement with the experiences of manual delineation. The aforementioned results demonstrate that SAM's performance in radiotherapy auto-delineation mirrors the clinical experience of human delineation, considering the variation across different sites and OARs. Importantly, SAM's auto-segmentation is achieved by only one pre-trained model, suggesting that SAM exhibits robust generalizability for automatic delineation regarding different sites in radiotherapy. 

Upon the inclusion of the box prompt, SAM's performance in auto-segmentation for radiotherapy showed further improvement. For most organs within the four reference sites, Dice and JAC scores have risen by 0.1 to 0.5, and previously unrecognized OARs such as the esophagus and parotid could be identified. This suggests that the box prompt, overall, is effective in enhancing SAM's performance in radiotherapy segmentation, and future research could consider employing diverse prompt methods for further improvement. However, some OARs with vague boundaries, such as large bowel, small bowel and cochlea, were still not well recognized. This limitation is true for all auto-segmentation algorithms based on CTs. SAM is sensitive to the clarity of the OAR boundaries in radiotherapy delineation images; for some OARs with unclear boundaries, multi-modality images such as MRI could be considered to assist in the segmentation of some OARs. In addition, 3D segmentation is required for clinical radiotherapy, which warrants further research.

Given the existing results, the SAM model demonstrates good generalization capabilities consistent with manual radiotherapy delineation, with a single model capable of executing all included test delineations for radiotherapy, and providing an effective prompt technology path for enhancing the performance of radiotherapy delineation. Judging from the Dice scores, SAM is able to meet the accuracy in the majority of OARs segmentation tasks required for clinical radiotherapy with Dice score higher than 0.7 when compared with manual delineation \cite{jameson2010review,vinod2016uncertainties}. Though parts of the OARs' segmentation results from SAM are lower than 0.7,  considering that the primary training of the SAM model largely relies on daily computer vision data, rather than professional clinical imaging data, the results are still acceptable, if not quite impressive \cite{Kirillov23}. This also suggests that future work can further fine-tune the SAM model in radiotherapy auto-segmentation through the incorporation of more clinical imaging data.

\subsection{Future Perspectives}

\subsubsection{Integration into Clinical Workflow}

SAM can play a pivotal role in streamlining the clinical workflow in radiation oncology. For example, in contouring where experts delineate the target and OARs manually, SAM can assist by providing preliminary segmentation. This could significantly reduce the burden on clinical staff, leaving them more time to focus on complex cases. Moreover, it could aid in minimizing inter-observer variability, a prevalent issue in manual contouring \cite{vinod2016review,lappas2022inter}.

\subsubsection{Support for Different Imaging Modalities}

SAM's adaptability across different imaging modalities can be further enhanced. The adaptation can be driven towards less commonly used but clinically relevant modalities such as Cone Beam CT (CBCT) \cite{posiewnik2019review} used for patient positioning during radiotherapy or PET/CT, which provides metabolic information in addition to structural anatomy \cite{seemann2004pet}.

\subsubsection{Collaborative AI-Human Decision Making}

SAM can act as a decision support system, not just providing initial segmentation but also flagging complex cases that require more in-depth scrutiny by the clinicians. This dual workflow could optimize the clinician's efforts towards cases that are more critical, improving overall treatment efficacy.


\subsubsection{Automatic Segmentation of OARs}

SAM's use can be expanded towards challenging tasks like delineating OARs with similar densities as the targets, such as the rectum in prostate cancer cases, or parotid glands in head \& neck cancers. With enough training on these specific sites, SAM can become a crucial tool in minimizing radiation-induced complications \cite{bhangoo2020acute,unkelbach2018robust}.

\subsubsection{Inter-Patient Adaptation}

SAM can also be tailored to aid inter-patient adaptation. Its ability to automatically adjust segmentation masks from one patient to another, factoring in individual anatomy and pathology, can be utilized to predict how a tumor might evolve or migrate over time. It could thus help in adapting the treatment plan for more personalized and precision radiation therapy.


\subsubsection{Custom Training for Specific Tasks}

A promising method to enhance SAM's performance for complex OARs involves task-specific fine-tuning. This could be realized by training SAM with an enriched dataset, containing a wide variety of examples of these complex structures.

In addition to leveraging the power of rich datasets, a novel method to optimize SAM's performance is through the use of adapter modules \cite{wu2023medical}. These adapters serve as lightweight components that are inserted at specific locations within the SAM architecture. The adapters are specifically tailored for a given task without requiring to fine-tune the entire model.

For radiation oncology tasks, these adapters can be designed and trained specifically for challenging tasks such as the segmentation of lymph nodes or vascular structures. They can also be implemented to handle ambiguity in the delineation of these complex structures. Rather than fine-tuning the entire model on a new dataset, these adapter modules allow the underlying pre-trained SAM parameters to remain frozen while the task-specific adapters are optimized.

Another critical feature of these adapters is their ability to operate in 3D space, providing the capability to process 3D medical images, a common data format in radiation oncology \cite{samarasinghe2021deep}. This would enable the model to better capture the spatial and depth correlations within the data, a crucial factor for accurate delineation of complex structures.

The adapter modules can be trained using a variety of self-supervised learning methods, such as contrastive embedding-mixup and shuffled embedding prediction \cite{tamkin2021dabs,wu2023medical,wantlin2023benchmd}, thereby enhancing their ability to handle complex segmentation tasks. These techniques may provide a robust solution to improve SAM's performance on complex OAR segmentation, thereby potentially revolutionizing its utility in radiation oncology.

\subsubsection{Enhancement with Language Prompts and LLM Encoder}

Another promising direction for enhancing SAM involves integrating it with language prompts and language models \cite{radford2021learning,liu2022survey,singhal2022large}, especially in application scenarios where domain-specific language and knowledge are beneficial \cite{cai2022coarse,li2023artificial,liu2023context,liao2023mask,rezayi2022clinicalradiobert,rezayi2022agribert}. Language prompts can be leveraged to guide the segmentation process, providing context or highlighting areas of interest in the images. In such a setup, the LLM encoder can convert the language prompts into a meaningful context vector that guides SAM's segmentation process. By aligning the language and vision model training through a shared context representation \cite{radford2021learning}, SAM may better delineate complex structures and ambiguous regions. This approach capitalizes on the strength of LLMs to understand and generate nuanced human language, potentially enabling more precise and clinically relevant segmentation.

\subsubsection{Reinforcement Learning and Expert Feedback}

Reinforcement learning \cite{sutton2018reinforcement}, using expert manual contouring data, can play a central role in the iterative improvement and validation of SAM. Expert manual contours serve as a gold standard in this context, with the differences between SAM's outputs and these expert contours acting as a source of feedback to guide the model's learning.

In reinforcement learning paradigms \cite{sutton2018reinforcement,zhou2021deep}, the system learns by interacting with its environment (in this case, the imaging data), and improves based on feedback (the differences between SAM's segmentation and expert-produced contours). SAM's "actions," which in this context are the segmentation it produces, would be continuously adjusted based on "rewards," or the degree of similarity to expert-produced contours. This could lead to a system where SAM progressively adapts its segmentation approach to match expert-produced contours more closely, with its progress validated against the expert-produced contours.

This iterative process, grounded in expert feedback, could continually refine SAM's capabilities, enhancing its precision and personalizing its approach over time. Importantly, reinforcement learning emphasizes learning from each segmentation task, turning every interaction with the data into a potential learning opportunity.

\subsubsection{Uncertainty Estimation}
Implementing a measure of uncertainty in SAM's output could be achieved by integrating Bayesian approaches into the learning process, such as Bayesian Convolutional Neural Networks \cite{gal2015bayesian}. These provide an estimate of the model's confidence in its predictions, allowing clinicians to identify areas where manual review and edits might be necessary.

\subsubsection{Integration of Clinical Knowledge}

SAM could be improved by integrating clinical information, such as the relationship between tumor location, size, and the expected effect of radiation. Machine learning models such as graph neural networks \cite{micheli2009neural} could be used to encode this information, where nodes could represent different anatomical sites, and edges could represent spatial relationships or common patterns of metastasis. Combining this with SAM's powerful image segmentation capabilities could yield a model that is both accurate and clinically relevant.

\section{Conclusion}
In this study, we explored the application of the SAM for medical image segmentation in radiation oncology. We found that SAM shows promising potential for aiding treatment planning, but it requires significant customization and refinement for clinical use. Our evaluation highlighted SAM's capacity to delineate large, distinct organs effectively, but revealed difficulties in segmenting smaller, intricate structures, especially with ambiguous prompts. This variability in performance across different anatomical sites and imaging modalities underlines the necessity to tailor SAM and similar foundation models to the specific requirements in medical domains.

Despite the opportunities presented by SAM, there is much room for improvement. Future developments should focus on refining SAM for specific clinical tasks, incorporating methods for uncertainty estimation, and integrating clinical knowledge into the model. Our study emphasizes that foundation models like SAM should augment, not replace, human expertise, and a balanced approach recognizing both the opportunities and limitations of large foundation models is vital. We foresee a bright future where incremental progress, guided by a harmonious integration of human and AI, propels advancements in radiotherapy.

\bibliography{LLM_refs}
\bibliographystyle{unsrt}

\end{document}